\documentclass{aa}
\usepackage{amssymb}
\usepackage{amsmath}
\usepackage{graphicx}
\usepackage{color,xcolor}
\usepackage[varg]{txfonts}
\usepackage{natbib}

\usepackage{placeins}
\usepackage{stfloats} 

\usepackage[colorlinks=true,citecolor=blue,urlcolor=blue,linkcolor=blue]{hyperref}
\usepackage[utf8]{inputenc}
\def\gr{$\gamma$-ray}

\begin{document}

\title{Modeling the propagation of very-high-energy $\gamma$-rays with the CRbeam code:\ Comparison with CRPropa and ELMAG codes}
\author{O.Kalashev$^{1}$, A.Korochkin$^{2}$, A.Neronov$^{3,4}$ and D.Semikoz$^{3}$}
\institute{
https://orcid.org/0000-0002-7982-1842
\and Université Libre de Bruxelles, CP225 Boulevard du Triomphe, 1050 Brussels, Belgium
\and Université de Paris, CNRS, Astroparticule et Cosmologie, F-75013 Paris, France
\and Laboratory of Astrophysics, Ecole Polytechnique Federale de Lausanne, CH-1015, Lausanne, Switzerland
}
\authorrunning{O.Kalashev et al}
\titlerunning{CRbeam}

\abstract
{
Very-high-energy  \gr s produce electron positron pairs in interactions with low-energy photons of extragalactic background light during propagation through the intergalactic medium. The electron-positron pairs generate secondary \gr s detectable by \gr\ telescopes. This secondary emission can be used to detect intergalactic magnetic fields (IGMF) in the voids of large-scale structure.
}
{
A new \gr\ observatory, namely, Cherenkov Telescope Array (CTA), will provide an increase in sensitivity for detections of these secondary \gr\ emission and enable the measurement of its properties for sources at cosmological distances. The interpretation of the CTA data, including the detection of IGMF and study of its properties and origins, will require precision modeling of the primary and secondary \gr\ fluxes.   
}
{
We asses the precision of the modeling of the secondary \gr\ emission using model calculations with publicly available Monte-Carlo codes CRPropa and ELMAG and compare their predictions with theoretical expectations and with model calculations of a newly developed CRbeam code.
}
{
We find that model predictions of different codes differ by up to 50\% for low-redshift sources, with discrepancies increasing up to order-of-magnitude level with the increasing source redshifts. We identify the origin of these discrepancies and demonstrate that after eliminating the inaccuracies found, the discrepancies between the three codes are reduced to 10\% when modeling nearby sources with $z\sim 0.1$. We argue that the new CRbeam code provides reliable predictions for the spectral, timing, and imaging properties of the secondary \gr\ signal for both nearby and distant sources with $z \sim 1$. Thus, it can be used to study gamma-ray sources and IGMF with a level of precision that is appropriate for the prospective CTA study of the effects of \gr\ propagation through the intergalactic medium.
}
{}

\maketitle

\section{Introduction}

The influence of the effect of pair production on the spectra of extragalactic sources of very-high-energy (VHE) \gr s has been well established via measurements of suppression of the highest energy \gr\ fluxes of blazars \citep{ebl_hess,ebl_veritas,ebl_magic}. Observations of this effect allow for the measurement of the spectrum of extragalactic background light (EBL), which is difficult to measure directly because of obscuration by Zodiacal light \citep{matsuura17}. The EBL is produced by cumulative emission from star-forming galaxies accumulated over their cosmological evolution \citep{franceschini08}. Its measurement over a range of redshifts provides information on the history of star formation in the Universe \citep{ebl_fermi}. The {\gr\ measurements of EBL  are also potentially interesting with regard to the search for narrow features in the EBL spectrum  \citep{Korochkin:2019pzr}, such as those produced by axion-like particles \citep{Korochkin:2019qpe}.

Absorption of the VHE \gr s on EBL results in production of  electron-positron pairs in the intergalactic medium. These pairs lose energy via inverse Compton scattering of the cosmic microwave  background (CMB) photons, thereby producing secondary \gr\ emission  that is detectable by \gr\ telescopes \citep{aharonian94}. The observational  visibility of this secondary emission depends on the strength and correlation length of magnetic field in the intergalactic medium  and on the energy range of the secondary \gr s. Very strong intergalactic magnetic field (IGMF) completely isotropises  trajectories of electrons and positrons so that the secondary emission appears as a \gr\ "halo" around a primary point source \citep{aharonian94}. Moderately strong IGMF induces extended secondary \gr\ emission, whose properties depend on the IGMF parameters \citep{neronov07,Murase:2008pe,neronov09}. Weak magnetic field results in the secondary flux morphologically indistinguishable from the point source, but identifiable through specific timing properties \citep{plaga95}. The non-detection of the secondary emission in the 1-100 GeV range has been used to derive a lower bound on IGMF at the level of $\sim 10^{-16}$~G under the assumption of large IGMF correlation length \citep{neronov10,2010MNRAS.406L..70T,2011MNRAS.414.3566T,2011ApJ...727L...4D,2012ApJ...744L...7T,2014ApJ...796...18A,HESS:2014kkl,Finke:2015ona,Tiede:2017aql,ackermann18,AlvesBatista:2020oio,Podlesnyi:2022ydu,MAGIC:2022piy}. 

The sensitivity of observations of the extended or delayed secondary emission from the electron-positron pairs in the intergalactic medium and of the measurements of IGMF will be crucially improved by the Cherenkov Telescope Array (CTA) \citep{2019scta.book.....C}. This will enable search for the extended secondary emission in a previously inaccessible energy range, thus extending the range of IGMF strengths that can be probed by the \gr\ technique \citep{korochkin21}.  The sensitivity improvement will result in a more precise characterization of the effect of absorption of the primary \gr\ flux by the EBL and more precise measurements of the primary \gr\ source spectra, as well as in the extension of the measurements into larger redshift range \citep{CTA:2020hii}. Moreover, the \gr\ measurements can help distinguish primordial magnetic fields produced during inflation from the field originating from cosmological phase transitions, by constraining the correlation length of the field    \citep{Korochkin:2021vmi}. Additionally, the \gr\ technique is also sensitive to the effect of the baryonic feedback among the large-scale structure that creates magnetised bubbles around galaxies \citep{Bondarenko:2021fnn}.

Improved sensitivity for the secondary \gr\ flux from extragalactic sources provided by CTA has to be matched by the improved precision of modeling of this flux. Several numerical codes have been previously developed for such modeling, including CRPropa \citep{CRPropa} and ELMAG \citep{elmag}. These codes are based on Monte Carlo modeling of development of electromagnetic cascades in the intergalactic medium in the presence of magnetic fields \citep{elyiv09}. Details of the implementation (of magnetic field structure, of cosmological evolution, etc.) differ between the codes. This may lead to code-dependent discrepancies in model predictions of the properties of secondary emission. Each code adopts certain models of the relevant physical processes and Monte-Carlo event generators. This results in systematic modeling errors that are difficult to assess.  

In the following, we compare the precision of the predictions of the CRPropa and ELMAG with those of the newly developed CRbeam code\footnote{CRbeam code is available at the link \url{https://github.com/okolo/mcray/tree/main/src/app/crbeam}} (first version of the code has been reported by \citet{Berezinsky:2016feh}). The modular structure of the \mbox{CRbeam} and \mbox{CRPropa} allows all relevant processes to be tested independently of each other. Specifically, we consider the Breit-Wheeler pair production and inverse Compton scattering on the EBL and CMB. For each interaction, we compare interaction rate and energy distribution of secondary particles inferred from simulations with the theoretical predictions. Also, disabling all interactions makes it possible to compare the propagation of electrons in a magnetic field. For ELMAG, on the contrary, such independent testing of interactions is impossible, therefore we use the results of simulations with ELMAG to compare the properties of the cascade signal, when all relevant interactions are turned on.

We find that the predictions for the spectral, imaging, and timing properties of the secondary emission may differ by about 50\% in the CTA's energy range of interest. We trace the origin of some of these discrepancies and find that they can be reduced after corrections. In particular, for relatively nearby sources with a redshifts $z\lesssim0.15$, the discrepancies between all three codes (CRbeam, CRPropa, and ELMAG) are down to 10\% in all our tests after introducing corrections. We find, however, that the scatter of the predictions increases substantially with increasing redshift even when our corrections are applied.  We argue that the main reason for the remaining discrepancies is the implementation of cosmological evolution of the EBL in CRPropa, which still lead to non-negligible differences between model calculations for sources at large redshifts.

We note that a certain fraction of the electron-positron energy may be carried away via plasma instabilities and thus affect the electromagnetic cascades \citet{2012ApJ...752...22B}. However, at the moment, there is no self-consistent description of the effect and its impact on the development of the cascades depends strongly on the assumptions used \citet{2013ApJ...770...54M,AlvesBatista:2019ipr, Alawashra:2022all}. For this reason, we do not include plasma instabilities in our comparison.

In all analyses presented here, we used the following versions of the codes: CRbeam 1.0, CRPropa3-3.1.7, and \mbox{ELMAG~3.01}. Additionally, we have verified that the results of \mbox{CRPropa} are stable in relation to previous versions \mbox{CRPropa3-3.1.5} and \mbox{CRPropa3-3.1.6}. 

For the cosmological parameters, we adopted the following values: Hubble constant $H_0=70$ km/s/Mpc, the matter density parameter
$\Omega_\mathrm{m} = 0.3,$ and the dark energy density parameter $\Omega_\mathrm{\Lambda} = 0.7$. We used these values in all three considered codes and they are remain unaltered in all tests.

\section{Absorption of the primary \gr s}
\label{sec:absorption}
We started our comparison from the modeling of absorption of primary \gr s on EBL. To date, the EBL spectrum and its evolution with redshift have not been measured with sufficient accuracy. Most of the knowledge on EBL based on EBL models  presented in the literature. Among them, we chose the EBL model of \citet{franceschini08} as the baseline model for our tests, since it is available in all codes and also its spectrum as a function of redshift is presented in tabular form.

\begin{figure*}
    \includegraphics[width=\linewidth]{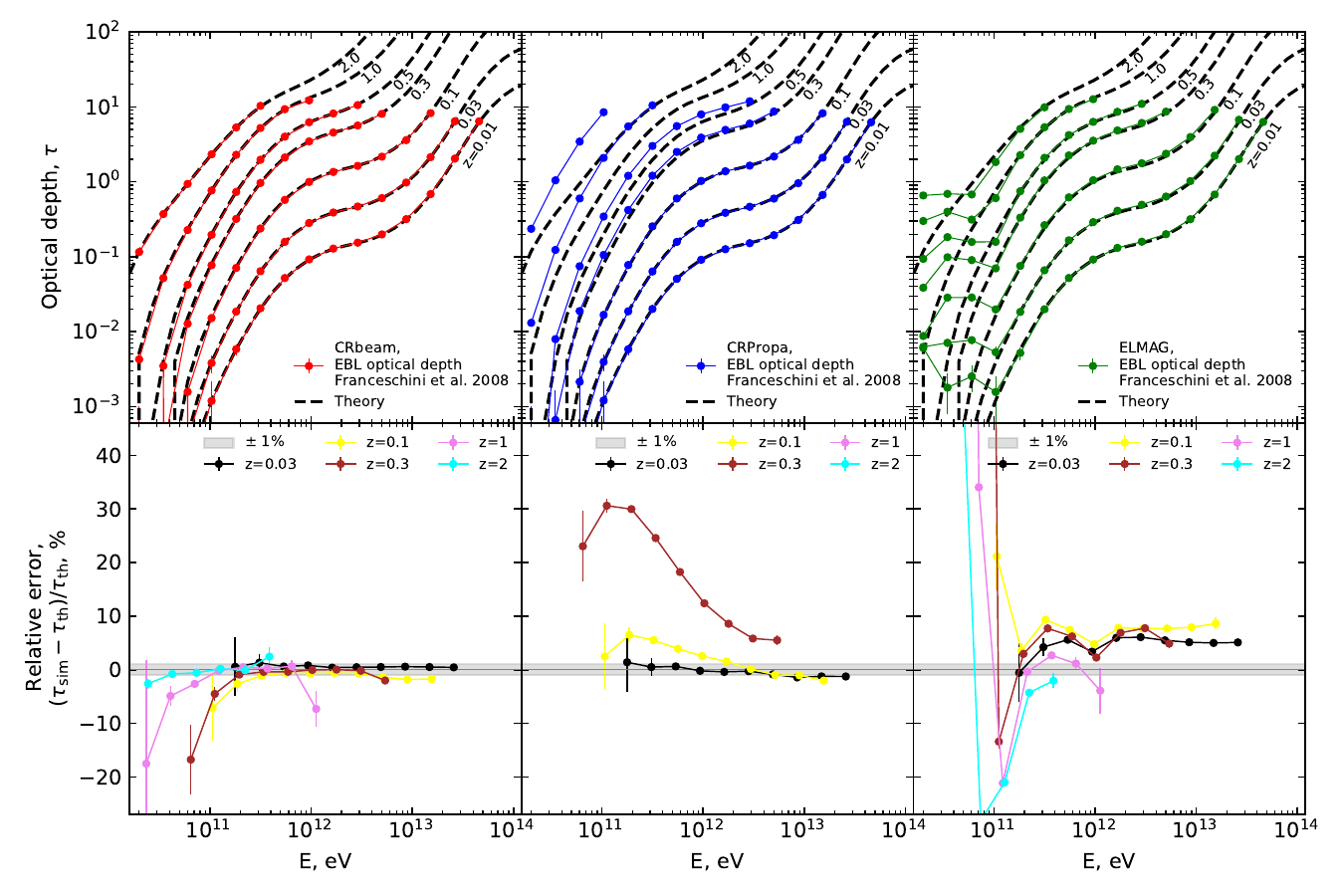}
    \caption{Comparison of the optical depth for primary \gr s for sources at cosmological redshifts in CRbeam, CRPropa and ELMAG. In all cases, the EBL model of \citet{franceschini08} is assumed. Top panels show the optical depth as a function of \gr\ energy for a range of redshifts. Bottom panels show discrepancies between the code calculations and the analytical model. For CRPropa, the differences between the analytical model and code calculations are not shown for redshifts of $>0.3$ because they exceed 50\%.
    \label{fig:EBL_Franceschini08_opdep}}
\end{figure*}

First, we calculated the optical depth by numerical integration of the EBL spectrum, taking into account its evolution over time. This corresponds to the black dashed line in the Fig.~\ref{fig:EBL_Franceschini08_opdep}. Next,  we calculated the optical depth using Monte Carlo codes, propagating $10^6$ monoenergetic photons and counting the number of survivors. The procedure is repeated for an array of observed photon energies and initial redshifts. We note, however, that deviations at the percentage level in the EBL optical depth may naturally arise due to different interpolation methods used in different codes. We do not consider differences of this kind in our analysis.

\begin{figure*}
    \includegraphics[width=\linewidth]{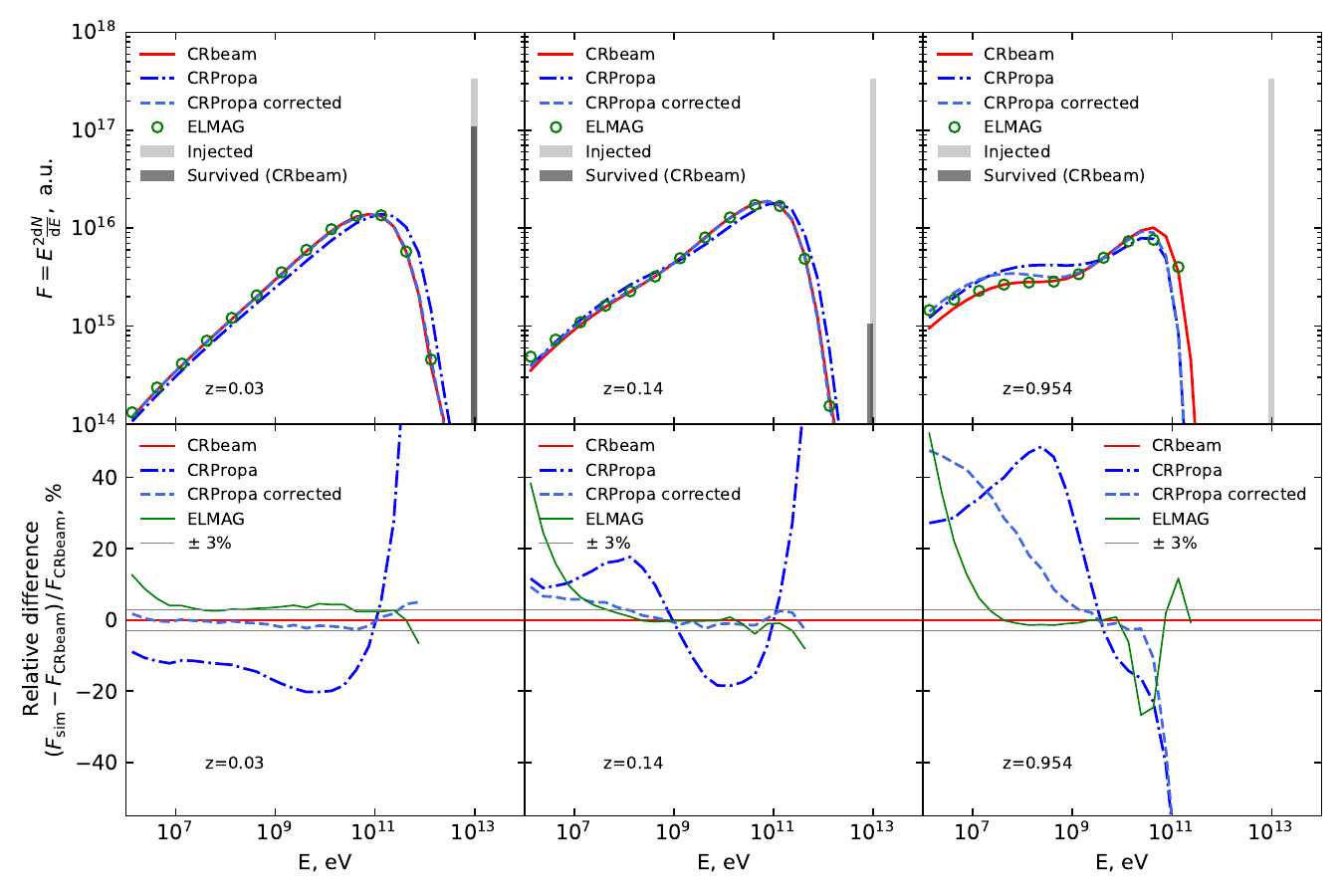}
    \caption{Differences in model predictions for the spectra of secondary \gr s produced by interaction of monoenergetic primary \gr\ beam with energy $E_{\gamma_0}=10$~TeV. Top panels show the primary and secondary \gr\ spectra for sources at different redshifts. Bottom panels show the differences between CRbeam, CRPropa, and ELMAG models. \label{fig:spec_mono10TeV}}
\end{figure*}

Figure~\ref{fig:EBL_Franceschini08_opdep} shows a comparison of the calculated optical depths for photons of different energies (\citet{franceschini08} EBL model). We can see that CRbeam modeling reproduces the assumed analytical model with percent-level precision in the energy range above $200$~GeV for a wide range of source redshifts. Differences of up to 20\% at the energies of $E<100$~GeV do not have substantial effects on the model results because the optical depth in this energy range is small, even for sources at redshift $z\sim 1$. 

The results of simulations with ELMAG are also in good agreement with theory in the energy region above 100 GeV, although it predicts $\sim5\%$ larger optical depth. Strong differences from theory are clearly noticeable in the energy range 10-100 GeV, where ELMAG predicts a significantly higher optical depth. For sources with redshifts $z\lesssim 0.5$, this discrepancy will have no effect, since in this case, even the wrong optical depth is much less than unity. Nevertheless, differences may arise for sources with $z\sim 1$, since in this case, in the ELMAG the optical depth $\tau\sim 1$ -- and not less than unity as it should be according to the theory. \footnote{While our manuscript was under revision, the new version \mbox{ELMAG}~3.03 was released in which our comments were taken into account and the bug with enhanced absorption at an energy $\sim$ 100 GeV was fixed. The simulations with ELMAG~3.03 are presented in Sect.~\ref{sec:appendixA}}

To the contrary, the calculations of CRPropa differ significantly from the assumed analytical model already for sources at redshifts $z \gtrsim 0.3$. The difference reaches 30\% at 100~GeV for the source at redshift $z=0.3$ and reaches up to 100\% for further away sources (not shown in the bottom panel of the figure). The discrepancy between the numerical and analytical model in the CRPropa code is due to simplified evolution of the EBL with redshift in which the real evolution of EBL is replaced by the renormalization of EBL spectrum taken at $z=0$ with a factor $s(z)$ \citep{CRPropa}. One can see from \citep{CRPropa} that this approach reproduces correctly the change the of EBL photon density over time. On the over hand it completely ignores the relative evolution of two EBL bumps because the shape of the spectrum remains fixed. Since the photon number density in the infrared peak of the EBL is much higher than in optical this means that the renormalization of the spectrum traces only evolution of the infrared peak while evolution of the optical peak becomes unphysical and strongly coupled to the infrared peak. Thus, one can understand the reason for overestimated absorption in CRPropa. Absorption of TeV gamma-rays occurs mainly on optical background photons. At redshift $z=0$ in the EBL model of \citet{franceschini08} the two EBL peaks are of comparable strength while at $z=1$ the far infrared peak become much stronger than optical. Consequently applying the EBL spectrum shape at $z=0$ two compute optical depth at $z=1$ results in enhanced absorption at energies $\sim$ 100~GeV - 10~TeV. This CRPropa limitation is known \citep{CRPropa}. Our result shows explicitly that such a simplification starts to produce unphysical results already starting from moderate redshifts $\sim 0.3$.

In order to make our analysis more general, we carried out the same simulations with two other popular EBL models: \citet{2011MNRAS.410.2556D} and \citet{2012MNRAS.422.3189G}. The advantage of the former model is that it has well defined error bars and so it is possible to estimate if the uncertainties introduced by Monte Carlo codes fall within the EBL uncertainty region. Unfortunately, this model is not available in the CRbeam code, so we made our comparison using only two codes. The latter EBL model is present in all three codes. The results are shown in Figs.~\ref{fig:EBL_Dominguez2011} and \ref{fig:EBL_Gilmore2012},  confirming the general trends found for the \citet{franceschini08} EBL model. CRbeam reproduces optical depth in wide range of energies and redshifts with percent-level accuracy. ELMAG generally predicts 5\% - 10\% stronger absorption at all redshifts almost independent of energy. Finally, CRPropa while producing accurate results for low-redshift sources suffers from simplified EBL evolution at high redshifts and significantly overestimates optical depth. For the EBL model of \citet{2011MNRAS.410.2556D}, the optical depth at $z>0.3$ from CRPropa is far beyond the uncertainty region  (see Fig. \ref{fig:EBL_Dominguez2011}). We note, again, that the enhanced absorption for $z>0.3$ in all three EBL models in \mbox{CRPropa} is the consequence of the fact that the infrared EBL bumps of all these models are higher than optical at high redshifts.

On the other hand, we would like to point out that the EBL spectral energy distribution is not well known, especially at large redshifts. Recent EBL measurements with \gr\ absorption \citep{MAGIC:2019ozu} have an order-of-magnitude uncertainty at $z\sim 1$. In this sense, the deviations from EBL models, arising from a simplified CRPropa approach, are of same order as our ignorance of high redshift EBL spectrum. In the future, with improved CTA measurements of sources at high redshifts, it will become possible to construct an improved EBL model at different redshifts. Such a model has to be used in tandem with the full redshift evolution code that is similar to CRbeam.

\section{Emission of secondary \gr s via inverse Compton scattering}
\label{sec:emission}

The precision in modeling properties of the secondary \gr\ emission depends on the precision in the calculation of the optical depth of the primary \gr\ flux (discussed in the previous section), as well as of the energy distribution of the produced pairs and of the differential cross-section of inverse Compton scattering. 

An example calculation of secondary \gr\ spectrum by CRbeam, CRPropa, and ELMAG is shown in Fig.~\ref{fig:spec_mono10TeV} for a monoenergetic primary \gr\ source at different redshifts, corresponding to known blazar type active galactic nuclei Mrk 421 ($z=0.03$), 1ES 0229+200 ($z=0.14$) and PKS 0502+049 ($z=0.954$), covering a wide range of redshifts. The magnetic field was assumed to be zero. 

Although all three codes fulfills the law of conservation of energy we find significant discrepancies in cascade spectra. The discrepancies reach 50\% for CRbeam-CRPropa comparison and some 30\% for CRbeam-ELMAG, in the case of the source at the redshift $z\sim 1$. Smaller discrepancies, at the level of 20\% over a broad energy range up to 100~GeV, are found for smaller redshift sources. Nevertheless, it is also in this case that the discrepancies rise to $>50\%$ at the high-energy end of the secondary \gr\ spectrum. 

\begin{figure*}
    \includegraphics[width=\linewidth]{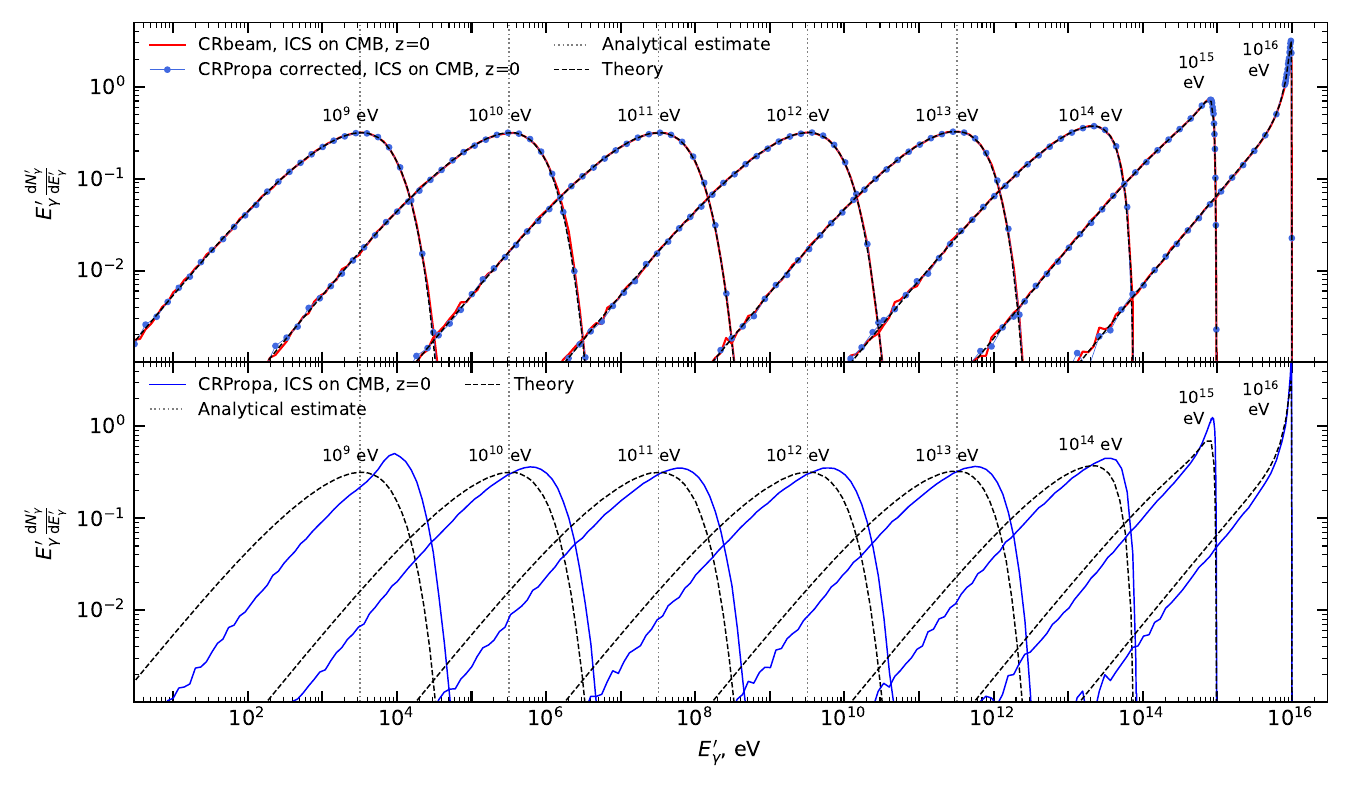}
    \includegraphics[width=\linewidth]{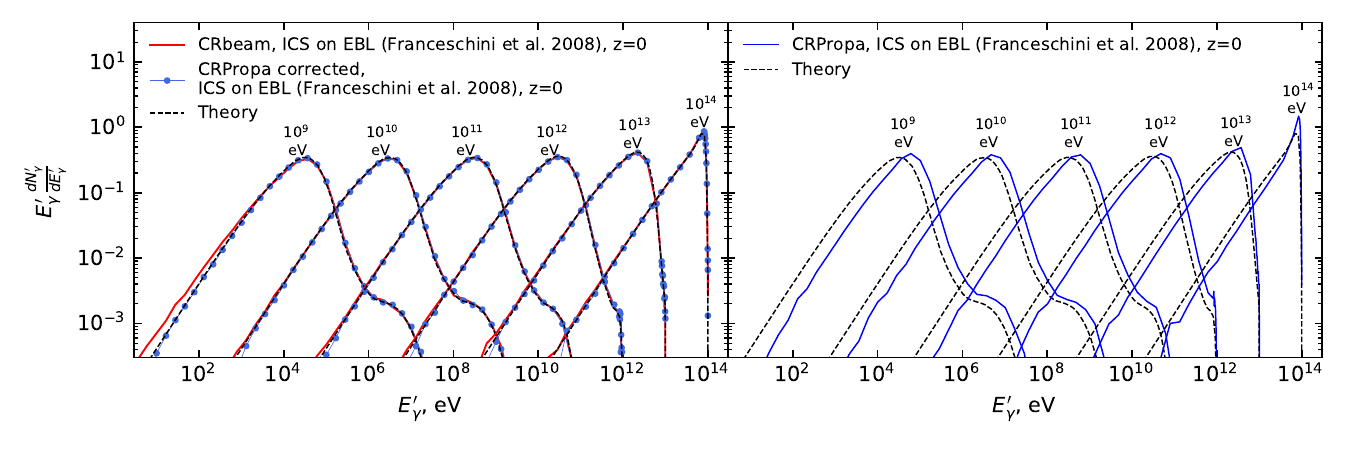}
    \includegraphics[width=\linewidth]{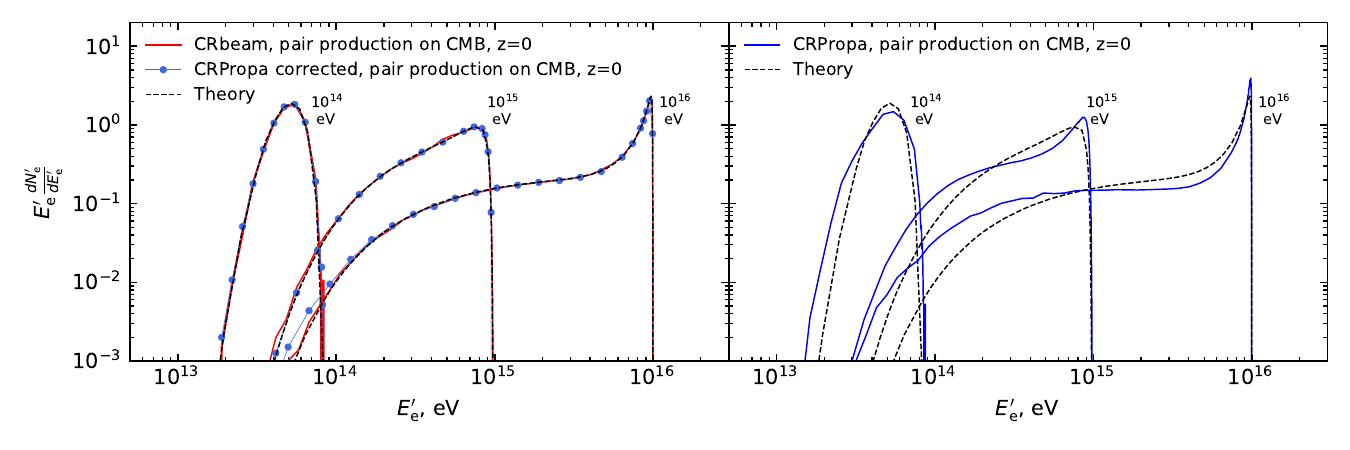}
    \caption{Energy distribution of \gr s produced via inverse Compton scattering on CMB (labeled as ICS on CMB) by electrons with different energies, indicated as tags for each set of spectra (top). 
    Spectra of \gr s from inverse Compton scattering on EBL (labeled as ICS on EBL) for a range of electron energies (middle). Energy spectra of pairs produced in interactions of \gr s with CMB photons (bottom). Energies of the primary \gr s are shown as tags for each group of spectra. In all cases, the codes used for the model calculations are specified in the legend.
    \label{fig:ICS}}
\end{figure*}
\begin{figure*}
    \includegraphics[width=\linewidth]{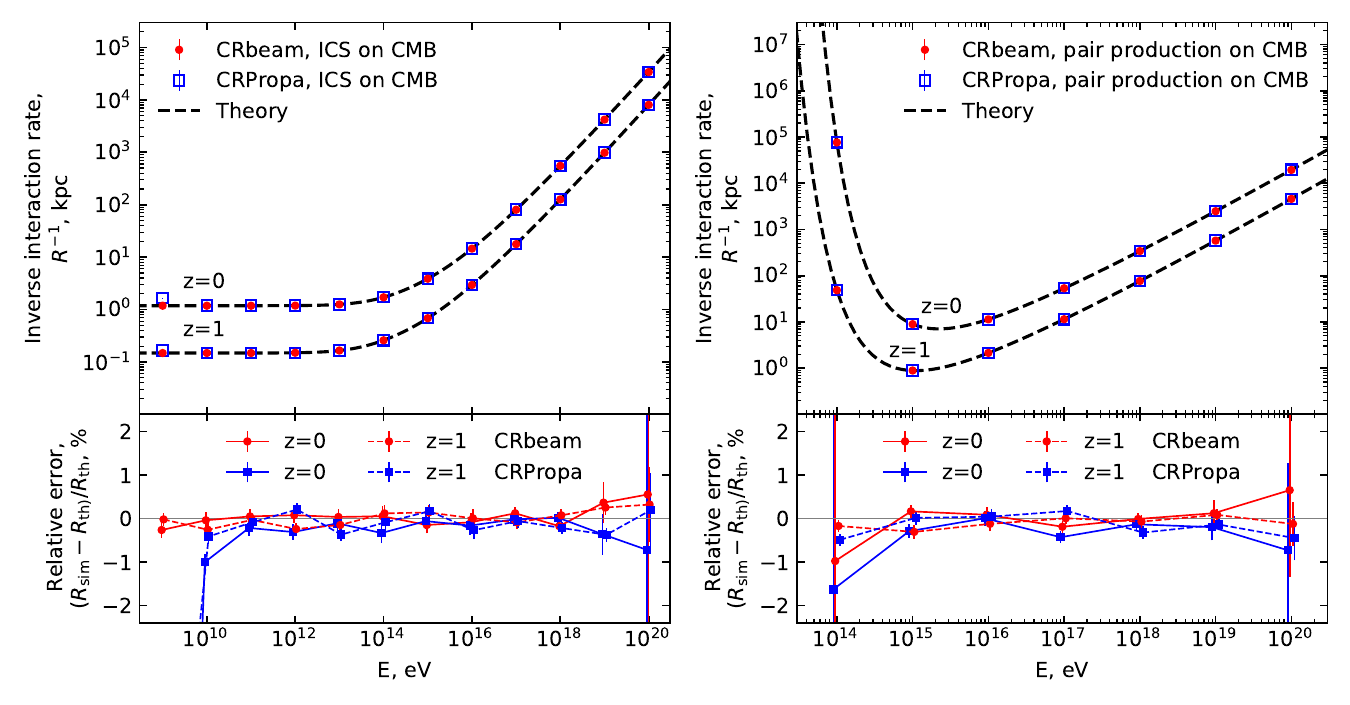}
    \caption{Interaction rates for the pair production and inverse Compton scattering on the CMB at the redshifts $z=0$ and $z=1$. 
    \label{fig:ICS_and_BW_onCMB_rate}}
\end{figure*}

The difference in cascade signal between CRbeam and \mbox{ELMAG} can be fully explained by the difference in the optical depth for the primary \gr s. Indeed, for a nearby source with a redshift $z=0.03$, a 5\% greater optical depth for 10 TeV gamma rays results in a 5\% amplification in the cascade signal over the entire energy range. For a more distant source with $z=0.14$, most of the primary gamma rays are absorbed; therefore, a 5\% difference in the optical depth does not appear and the normalization of the cascade signals in the region above 100 MeV coincide. However, increased absorption leads to an increase in the suppression of the high-energy part of the secondary signal. This is manifested in the fact that in the ELMAG the cascade signal is weaker in the region above 100 GeV and stronger in the region below 100 MeV. The same is true for a source at the redshift $z=0.954$. In addition, a dip at 10-100 GeV is noticeable in the ELMAG spectrum corresponds to the  order-of-magnitude excess absorption in ELMAG in this energy range, which leads to an even stronger amplified cascade signal below 100 MeV \footnote{Calculations with the updated version ELMAG~3.03 are shown in Figs.~\ref{fig:EBL_Franceschini08_opdep_Elmag303}~and~\ref{fig:spec_mono10TeV_Elmag303}}.

Unlike ELMAG, the difference between CRPropa and CRbeam cannot be explained by this single cause. Part of the discrepancy is certainly due to the the differences in the calculation of the optical depth for the primary \gr s producing pairs on the EBL. However, further differences in the modeling of the secondary flux are involved. This is indicated by the significant differences of the results of calculation with different codes for low-redshift sources, for which the optical depth calculations do not show strong discrepancies. 

We have found that for low-redshift sources, the main difference comes from the calculation of the inverse Compton scattering spectra by electrons and positrons. Figure~\ref{fig:ICS} shows a comparison of analytically calculated spectra of the inverse Compton scattering on the CMB with the output of CRbeam and CRPropa for this process. We can see that the CRPropa calculations do not match the theoretical formula \citet{RevModPhys.42.237}. The discrepancies are of the order-of-one all over the energy range of the inverse Compton emission for electrons with energies up to $10^{16}$~eV. This large discrepancy is "smeared" and becomes less noticeable in the calculation of the spectrum of inverse Compton emission from a broad energy distribution of electrons  (shown in Fig.~\ref{fig:spec_mono10TeV}). Vertical dashed lines in the Fig.~\ref{fig:ICS} corresponds to the mean energy of secondary gamma-rays after inverse Compton scattering $\langle E'\rangle=4/3 \gamma^2 \langle E_\mathrm{CMB}\rangle,$ where $\gamma$ is the gamma factor of the electron and $\langle E_\mathrm{CMB}\rangle=6.3\times10^{-4}$ eV is the mean energy of CMB photons. It is clear that the mean energy of the secondary \gr s in CRPropa is higher than theoretically expected.

We have also identified a similar problem of the CRPropa code in the calculation of the inverse Compton scattering on the EBL (see middle panel of Fig.~\ref{fig:ICS}) and in the spectra of $\gamma\gamma$ pair production. The bottom panel of Fig.~\ref{fig:ICS} illustrates this problem for the case of pair production on CMB. In CRPropa, the electron-positron spectrum differs from the known analytical formulae \citep{PhysRev.155.1404}, shown by the dashed lines. Generally, CRPropa predicts wider energy distribution of pairs than it should be according to the theory.

Having noticed these discrepancies, we analyzed the implementation of the CRPropa code and found out that the error with the energy distribution of secondary particles has a common origin for all three cases (inverse Compton scattering on CMB and EBL, and pair-production on CMB). The error occurs from the pre-calculated tables that come with the code. These tables are used to sample the energy at the center of mass as the interaction occurs. After recalculating the tables, the results obtained with the CRPropa with these corrected tables match analytical calculations, as can be seen from Fig.~\ref{fig:ICS} (in all plots, we use the "CRPropa-corrected" label for the CRPropa with recalculated tables). After this correction, model calculations of the spectra of secondary emission for our benchmark case of monoenergetic primary \gr\ beam agree between CRbeam and CRPropa for low-redshift sources down to $\le 5\%$ level, as seen in Fig.~\ref{fig:spec_mono10TeV} \footnote{We have informed the developers of CRPropa about this problem and while the paper was under review, a new version CRPropa3-3.2 was released \citet{AlvesBatista:2022vem}. In the new version the bug with pre-calculated tables (which affected all electromagnetic interactions) was fixed, so the results marked as 'CRPropa corrected' on our plots can be read as 'CRPropa3-3.2'}. The discrepancies still grow with increasing source redshift, for the reasons explained in the previous section.

\begin{figure*}
    \includegraphics[width=\linewidth]{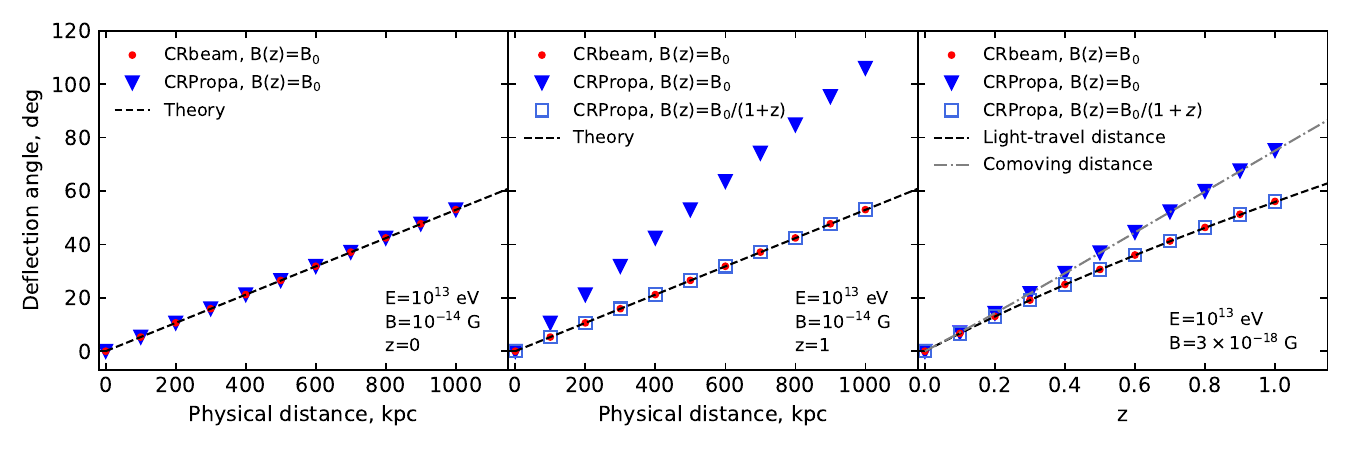}
    \caption{Deflection of electrons of different energies in constant magnetic field with fixed physical strength in CRbeam and CRPropa.
    \label{fig:propag_constB}}
\end{figure*}
\begin{figure*}
    \includegraphics[width=\linewidth]{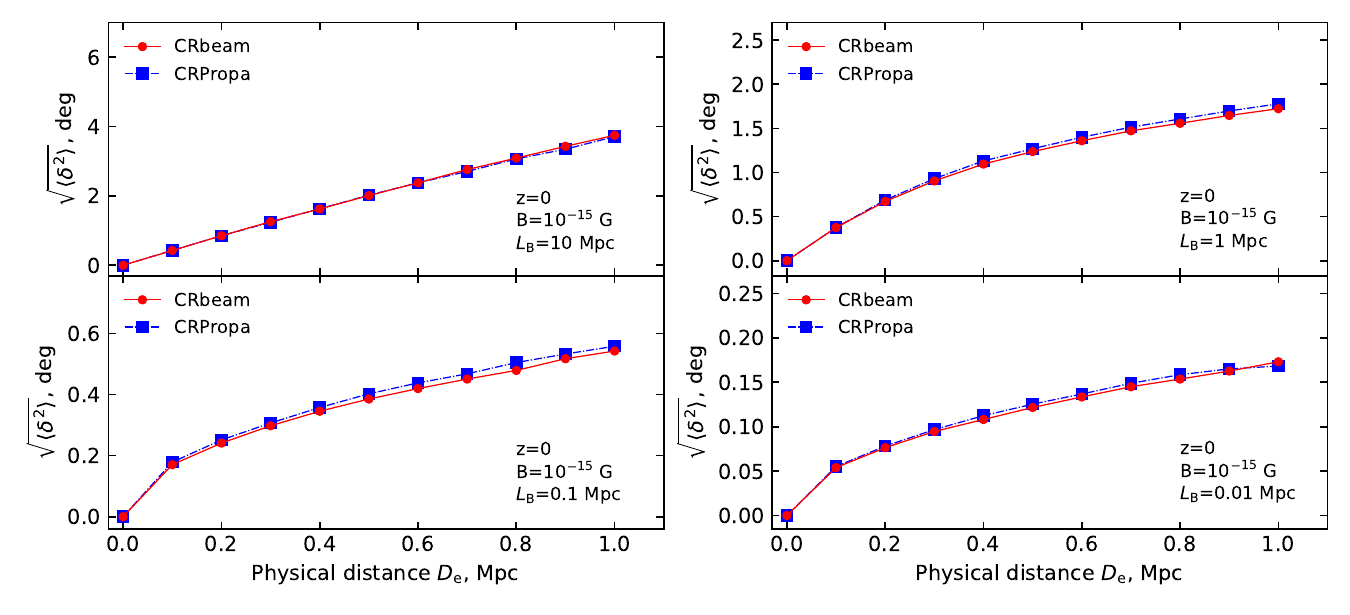}
    \caption{Deflections of electrons at $z=0$ in turbulent magnetic fields with different maximum scales $L_B$. Panels show the root mean square deflection angle dependence on the physical path length.
    \label{fig:propag_turbB}}
\end{figure*}

We also verified that the interaction rates for inverse Compton scattering and pair production on CMB calculated using CRbeam and CRPropa are in agreement with each other and with a relative difference from theory not exceeding one percent (see Fig.~\ref{fig:ICS_and_BW_onCMB_rate}). The 30\% lower inverse Compton scattering interaction rate in CRPropa for energies lower than 10 GeV is present only in the  CRPropa3-3.1.7 version and absent from previous versions.

Finally, we checked the distributions of secondary particles after electromagnetic interactions at redshift $z=1$. The conclusions are the same as in the case of $z=0$. The distributions from CRbeam perfectly follow the theoretical curves, while CRPropa results deviate from theory in the same way as at $z=0$. These deviations are also eliminated by recalculating interaction tables, as described above.

\section{Propagation of electrons and positrons}
\label{sec:propagation}
Imaging and timing properties of the secondary \gr\ signal are determined by the details of propagation of electrons and positrons under the influence of IGMF. The inverse-Compton scattered \gr s are emitted after electrons and positrons are deflected by an angle $\delta$ determined by the energy of electron, $E_e$, and strength of the magnetic field, $B$. In the simplest case of homogeneous magnetic field the angle is given by the analytical formula: 
\begin{equation}
    \delta =\frac{D}{R_L}=\frac{eB}{E_e}D
\end{equation}
where $e$ is the electron charge, $D$ is the propagation distance, and $R_L=E_e/eB$ is the gyroradius.

Both CRPropa and CRbeam correctly solve the equations of motions of the electrons in magnetic fields. However, the approaches of the two codes differ when accounting for cosmological effects. In CRbeam, the particle coordinates are expressed in physical distances, while in CRPropa, the comoving coordinates are used \citep{CRPropa}. This difference affects the propagation of electrons which can be illustrated by a simple test. Suppose the electron is deflected by a homogeneous magnetic field that is perpendicular to the electron's velocity. Figure~\ref{fig:propag_constB} shows the calculation of deflection angle of electrons for this simplest case, found with the CRbeam and CRPropa codes. We can see that both calculations give the same result at the redshift $z=0,$ where the physical and comoving distance elements are equal, left panel of Fig.\ref{fig:propag_constB}. However, the CRPropa calculation predicts $(1+z)$ stronger deflection at the redshift $z=1$, which is explained by the fact that $\mathrm{d}x_{\mathrm{comoving}}=(1+z)\,\mathrm{d}x_{\mathrm{physical}}$, so that an electron travels $(1+z)$ longer distance in the same magnetic field (see middle panel of Fig.~\ref{fig:propag_constB}).

This difference becomes most pronounced when the electron travels over cosmological distances. The right panel of Fig.~\ref{fig:propag_constB} shows the deflection of electron starting at redshift z (x-axis of the panel) and caught at $z=0$. We can see that the deflection angles in CRPropa are proportional to the comoving distances and in CRbeam to the light travel distances. The simplest way to get the correct deflection angles in CRPropa is to set the evolution of the magnetic field with redshift, namely, $B(z)=B_0/(1+z)$, what can be done using the built-in methods of CRPropa.

\begin{figure*}
    \includegraphics[width=\linewidth]{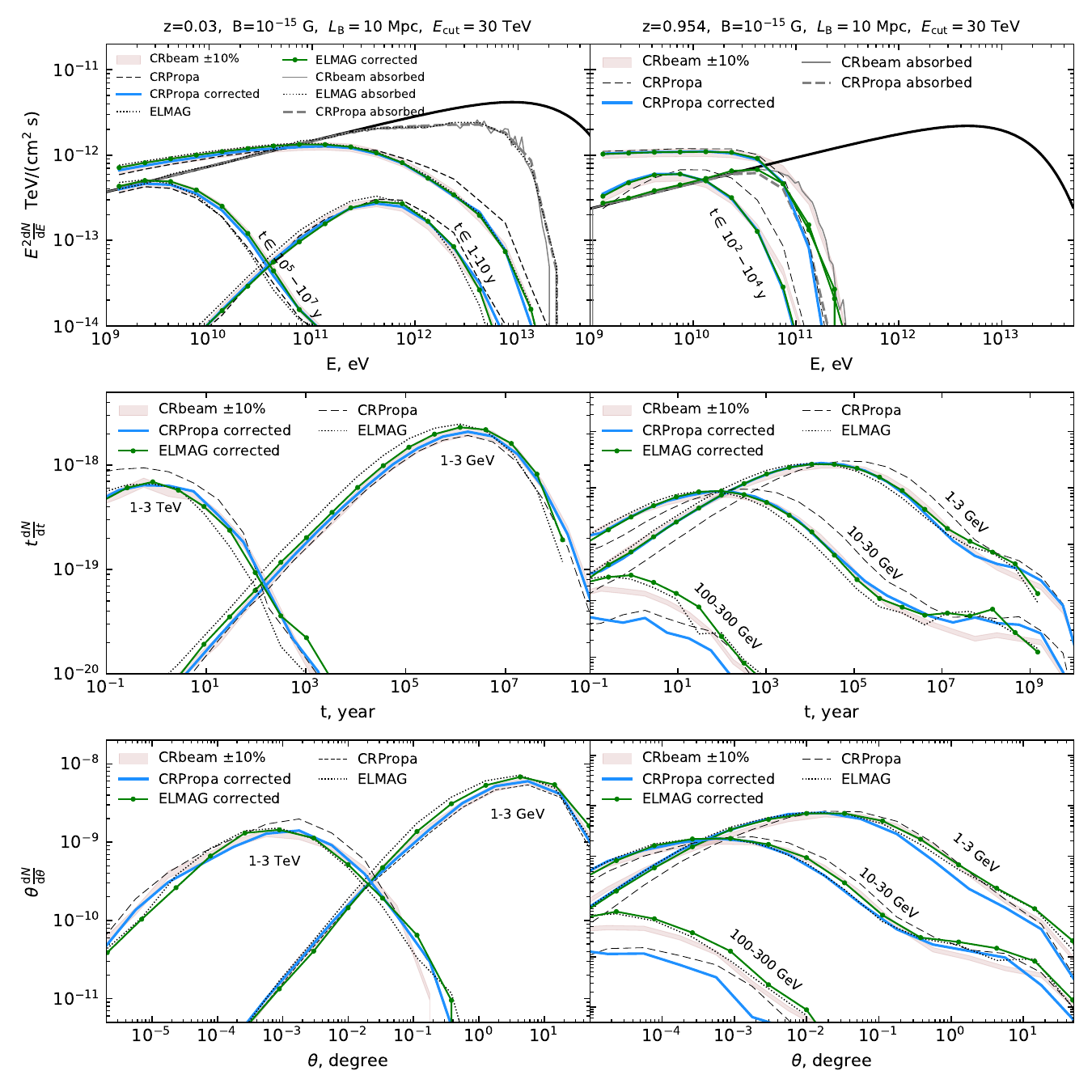}
    \caption{Spectral (top), timing (middle), and imaging (bottom) properties of primary and secondary \gr\ signals for a IGMF with strength $10^{-15}$~G and maximum scale $L_B=10$~Mpc. Left panels show calculations for the source at redshift $z=0.03$, right panels are for the source at redshift $z=0.954$. Calculations with CRbeam code (pink solid) compared with CRPropa (black thin solid), corrected CRPropa (blue solid), ELMAG (black dotted), and corrected ELMAG (green solid with dots) codes. The cascade signals on the upper panels are shown for the sources been active in the past in the time intervals indicated as tags near the curves while the curve without the tag corresponds to the permanently active source. The intrinsic spectrum of the source (thick black solid line) remains fixed during source activity. Note:\ on the plots of the angular and time distributions, the relative normalization of the curves for different energies is adjusted for visibility.
    \label{fig:FullCascade_z003_z0954}}
\end{figure*}

The field rescaling trick should also be used when an electron propagates in a realistic turbulent magnetic field with the strength $B_0$ and correlation length $\lambda_\mathrm{B0}$. The physical evolution with redshift of this field is given by the simple relations: ${B(z)=(1+z)^2\,B_0}$, ${\lambda_\mathrm{B} (z)=\lambda_\mathrm{B0}/(1+z)}$. However, because of the use of comoving coordinates in CRPropa, it is necessary to set ${B(z)=B_0\,(1+z)}$ to get correct deflection angles. This modified evolution should be used both when the correlation length of the field is greater than the propagation distance and when it is smaller.

We have compared the results on electron deflection in turbulent magnetic field at $z=0$. For the magnetic field we used Kolmogorov spectrum with the maximum scale $L_B$ and minimum scale equal to $L_B/100$. To ensure sufficient accuracy we limit maximum step size of the electron to $L_B/50$. Figure~\ref{fig:propag_turbB} shows the dependence of deflection angle on propagation distance for electrons moving in turbulent magnetic fields. The codes are in good agreement with each other and reproduce analytical formula of \citep{neronov09} in the case of a small correlation length:
\begin{equation}
    \sqrt{\left<\delta^2\right>}=\frac{\sqrt{D \lambda_B}}{R_L}=\sqrt{D\lambda_B}\frac{eB}{E_e}
,\end{equation}
where $\lambda_B$ is the magnetic field correlation length. For the Kolmogorov spectrum, $\lambda_B \approx L_B/5$.

\begin{figure*}
    \centering
    \includegraphics[height=0.9\textwidth]{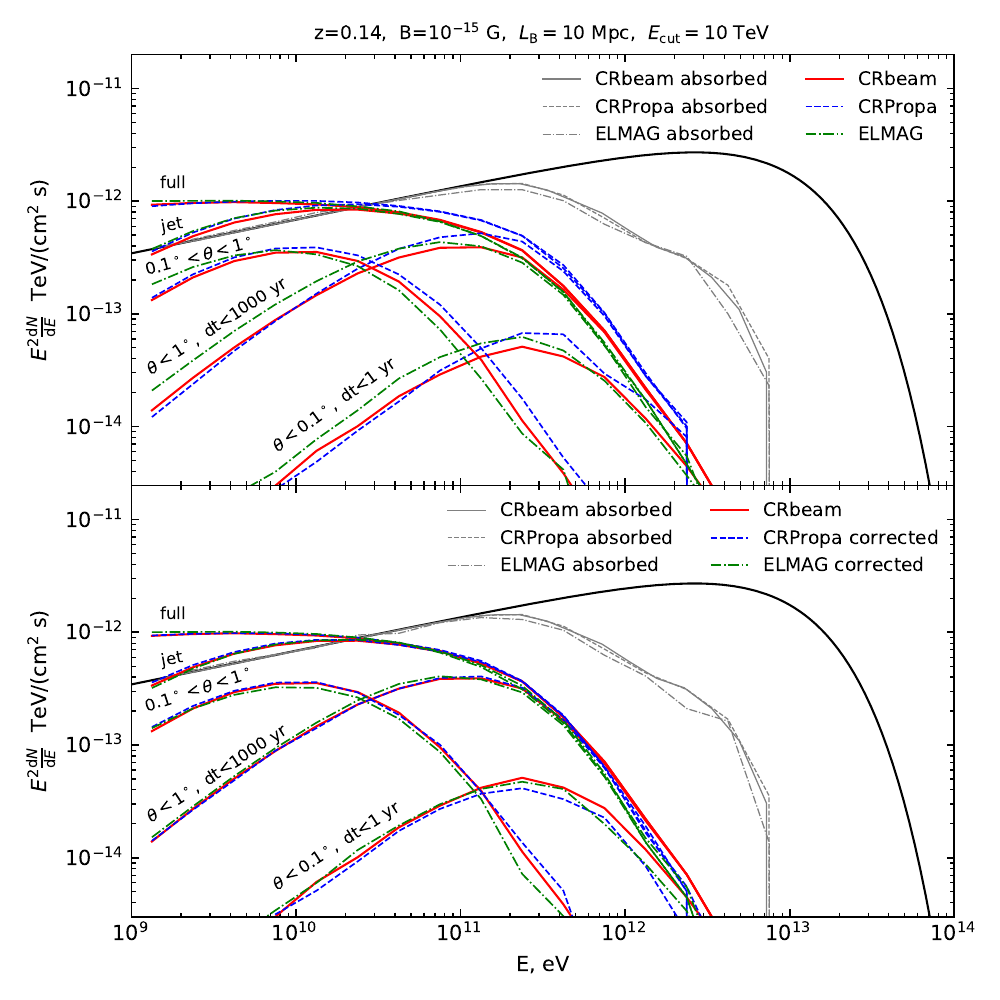}
    \caption{Cascade contributions for the source at the redshift $z=0.14$ and for a    turbulent IGMF with Kolmogorov spectrum with a strength of $B=10^{-15}$~G and maximum scale of $L_B=10$~Mpc (corresponding correlation length $\lambda_B \approx 2$~Mpc). The jet opening angle was assumed to be  $\theta_\mathrm{jet}=5^\circ$, and the observer was at an angle $\theta_\mathrm{obs}=3^\circ$ with respect to the jet axis. Red solid lines is for CRbeam code, blue dashed lines is for CRPropa code and green dash-dotted lines:\ the ELMAG code. Line labels indicate the constraints on the source activity time $\mathrm{d}t$ and angular distance from the source $\theta$ that were used when constructing the cascade signal. The label "jet" denotes the cascade signal without constraints on $\mathrm{d}t$ ($\mathrm{d}t=\infty$) and $\theta$ ($\theta=2\pi$). For comparison, we also added a cascade spectrum in the absence of a magnetic field (label "full"). 
    \label{fig:FullCascade_z014}}
\end{figure*}

We also tested the turbulent magnetic field generator in ELMAG. Using the built-in \texttt{test\_turbB} function, we separately calculated the average value of each component of the magnetic field. Assuming that the field of the strength $B$ is uniform and isotropic, it was expected that
\begin{equation}
    \sqrt{<B_x^2>} = \sqrt{<B_y^2>} = \sqrt{<B_z^2>} = B/\sqrt{3}
.\end{equation}
As a result of tests, we found that the values of the x and y components of the magnetic field are about 20 percent less than predicted theoretically, while there is no such difference for the z component. The reason for this was not found, so we simply re-scaled the x and y components of the field to the correct value. The results of ELMAG with such a re-scaled field are indicated on the plots as "ELMAG-corrected."

\section{Properties of primary and secondary emission from sources at different distances}

The realistic modeling of spectral, imaging, and timing properties of the primary and secondary \gr\ signals from extragalactic sources combines the modeling of the pair production, inverse Compton scattering, and electron propagation (discussed above). Discrepancies in the modeling of these ingredients propagate to the discrepancies of the overall signal models. We discuss these discrepancies in this section. As in the previous sections, here we use the EBL model of \citet{franceschini08}.

The left side of Fig.~\ref{fig:FullCascade_z003_z0954} shows the result of modeling of the primary and secondary \gr\ signals for the nearby blazars such as Mrk 421 and Mrk 501. For the intrinsic spectrum of the source, we assume a power law model, with a cutoff at $E_\mathrm{cut}=30$ TeV and a slope $\alpha=1.7$.

As discussed in Sect.~\ref{sec:absorption}, all the codes provide identical predictions for the optical depth for the primary \gr s in this case. As noticed in Sect.~\ref{sec:emission}, predictions for the spectra of pair production and inverse Compton scattering in the CRPropa code deviate from known analytical formulae. This leads to discrepancies between the models of secondary \gr\ fluxes in CRbeam and CRPropa codes, noticeable in the TeV energy range, which is most interesting for CTA. The top panel of the figure shows that the differences reach a factor of two above 3 TeV. The middle and bottom panels of the figure show the temporal and angular profiles of extended emission after an instantaneous injection of energy at the source at time $t=0$ with the spectrum described above. The differences in the spectral properties of the secondary signal between CRbeam and CRPropa are not so pronounced in the low-energy temporal and angular profiles in the lower energy bands, accessible to the Fermi Large Area Telescope (Fermi/LAT) instrument. However, these differences expand in the CTA energy band above 100~GeV. As discussed in Sect.~\ref{sec:emission}, the discrepancy between the CRbeam and CRPropa predictions can be removed after a correction to the CRPropa code. Difference between CRbeam and ELMAG also decreases after re-scaling magnetic field in ELMAG (see Sect.~\ref{sec:propagation}). In this case, all the codes produce results that agree at  the$\sim 10\%$ level, as can be seen from the left side of Fig.~\ref{fig:FullCascade_z003_z0954}.

To the contrary, corrections to the CRPropa code do not remove the discrepancies between different model predictions for a source at high redshift (for details, see the right side of Fig.~\ref{fig:FullCascade_z003_z0954}). A very significant difference in model predictions is noticeable at the highest energy ends of the spectra, visible in the top right panel of Fig.~\ref{fig:FullCascade_z003_z0954}. CRPropa predicts a strong suppression of the flux above 200~GeV, while CRbeam predicts a gradual suppression in the energy range between 200 and 300~GeV. This discrepancy is crucially important for the study of high-redshift sources with CTA. The discrepancy is noticeable also in the attenuated primary source spectra shown by the grey curves in Fig.~\ref{fig:FullCascade_z003_z0954}. We attribute this discrepancy for high-redshift sources in CRPropa to the simplified modeling of absorption on EBL, as described in Sect.~\ref{sec:absorption}. Such a modeling approach is insensitive to separate evolution of optical EBL peak and have to be improved before any sensible results on the EBL evolution can be extracted from the CTA data. 

The discrepancies in the timing and imaging properties of the secondary \gr\ signal are just as large, as shown in the middle and bottom panels on the right side of Fig.~\ref{fig:FullCascade_z003_z0954}. Contrary to the low-redshift sources, differences in predictions for the peaking time of the delayed signal are large already for lower energy \gr s detectable with Fermi/LAT. 

ELMAG results on high-redshift sources after correction are in good agreement with CRbeam, except for the largest time delays and deflection angles. The difference around 100 GeV can be explained by the increased absorption in ELMAG (see Sect.~\ref{sec:absorption}).

The results for the source 1ES 0229+200 at redshift $z=0.14$ are presented in  Fig.~\ref{fig:FullCascade_z014}. Calculations are done for IGMF with Kolmogorov spectrum with strength $B=10^{-15}$~G and maximum scale $L_B=10$~Mpc. In this figure the full cascade contribution for zero magnetic field is denoted as "full." The curves marked "jet" show the extended and delayed fluxes for non-zero IGMF for different angular and time intervals. The results for the \mbox{CRbeam} code are compared with CRPropa and ELMAG before (upper panel) and after (lower panel) corrections. All the codes give similar results for most of the parameters, after corrections.

\section{Discussion}
Our comparative study of model predictions for the primary and secondary very-high-energy \gr\ signals from astronomical sources at different redshifts has revealed significant differences in model predictions obtained with different Monte Carlo codes: CRbeam, CRPropa, and ELMAG. We have identified large differences in the model calculations of observables that are crucial for cosmological probes that will be enabled by the new CTA \gr\ observatory.

We  found the origin of these discrepancies and eliminated most of them. Once corrections are introduced to the codes, the discrepancies are significantly reduced. We contacted the developers of ELMAG and CRPropa and they fixed problems discussed in this paper in the new versions of their codes, ELMAG~3.03 and CRPropa3-3.2. 
Most remarkably, following these corrections, the difference in the cascade signal does not exceed 3\% between the codes when modeling a nearby source with $z\lesssim0.1$ and zero IGMF. Including nonzero IGMF slightly increases the scatter in the results, but  it is generally less than 10\%. This number can be considered as an estimate of the systematic modeling error in the measurements of cosmological evolution of EBL and measurements of IGMF with CTA.

The agreement of the results degrades as the redshift of the source approaches $z\sim1$. The largest discrepancies, which can reach up to an order of magnitude, are observed in the results of CRPropa, as compared with the results of CRbeam and ELMAG (which remain close). We attribute this divergence to the simplified modeling of the EBL evolution with redshift in CRPropa. Such a simplified approach assumes a fixed shape for the EBL spectrum, which limits the possibilities for studying its evolution by modeling the spectra of distant sources.

\begin{acknowledgements}
    The work of D.S. and A.N. has been supported in part by the French National Research Agency (ANR) grant ANR-19-CE31-0020.
\end{acknowledgements}

\bibliographystyle{aa}
\bibliography{references.bib}

\begin{appendix}

\section{ELMAG~3.03}\label{sec:appendixA}
\begin{figure*}[b]
    \begin{center}
        \includegraphics[width=0.4\linewidth]{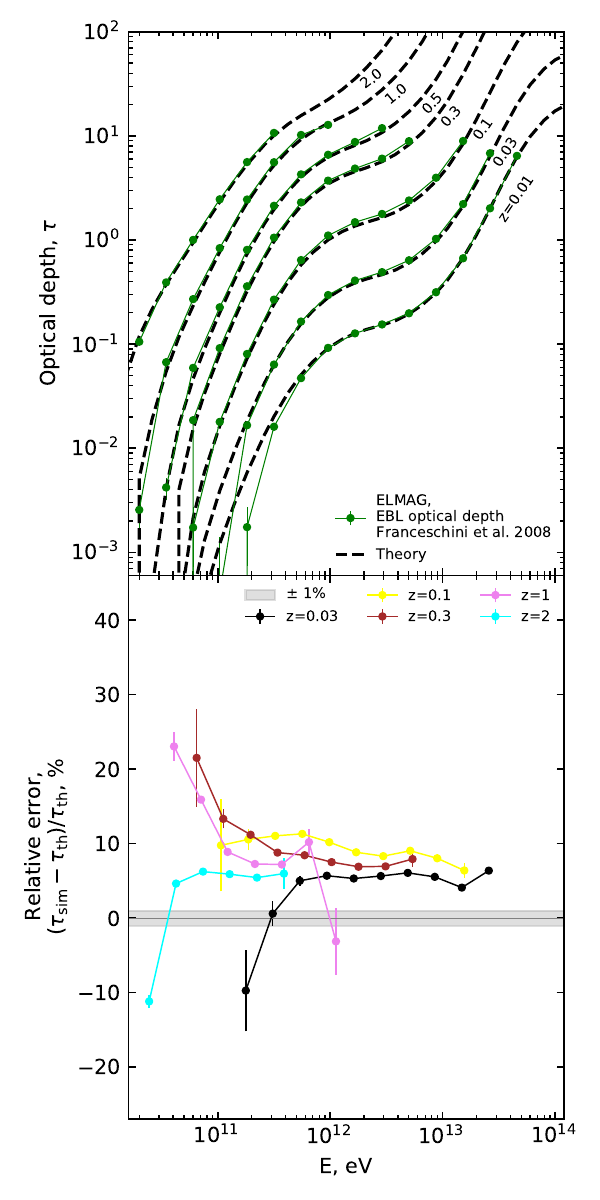}
        \caption{Comparison of the optical depth for primary \gr s for sources at cosmological redshifts calculated with ELMAG~3.03 for the EBL model of \citet{franceschini08}. Details are the same as in Fig.~\ref{fig:EBL_Franceschini08_opdep}. We can see the improvement in the energy range below 100 GeV compared with ELMAG~3.01, see Fig.~\ref{fig:EBL_Franceschini08_opdep}.\label{fig:EBL_Franceschini08_opdep_Elmag303}}
    \end{center}
    \vspace{5.5cm}
\end{figure*}
\begin{figure*}
    \includegraphics[width=\linewidth]{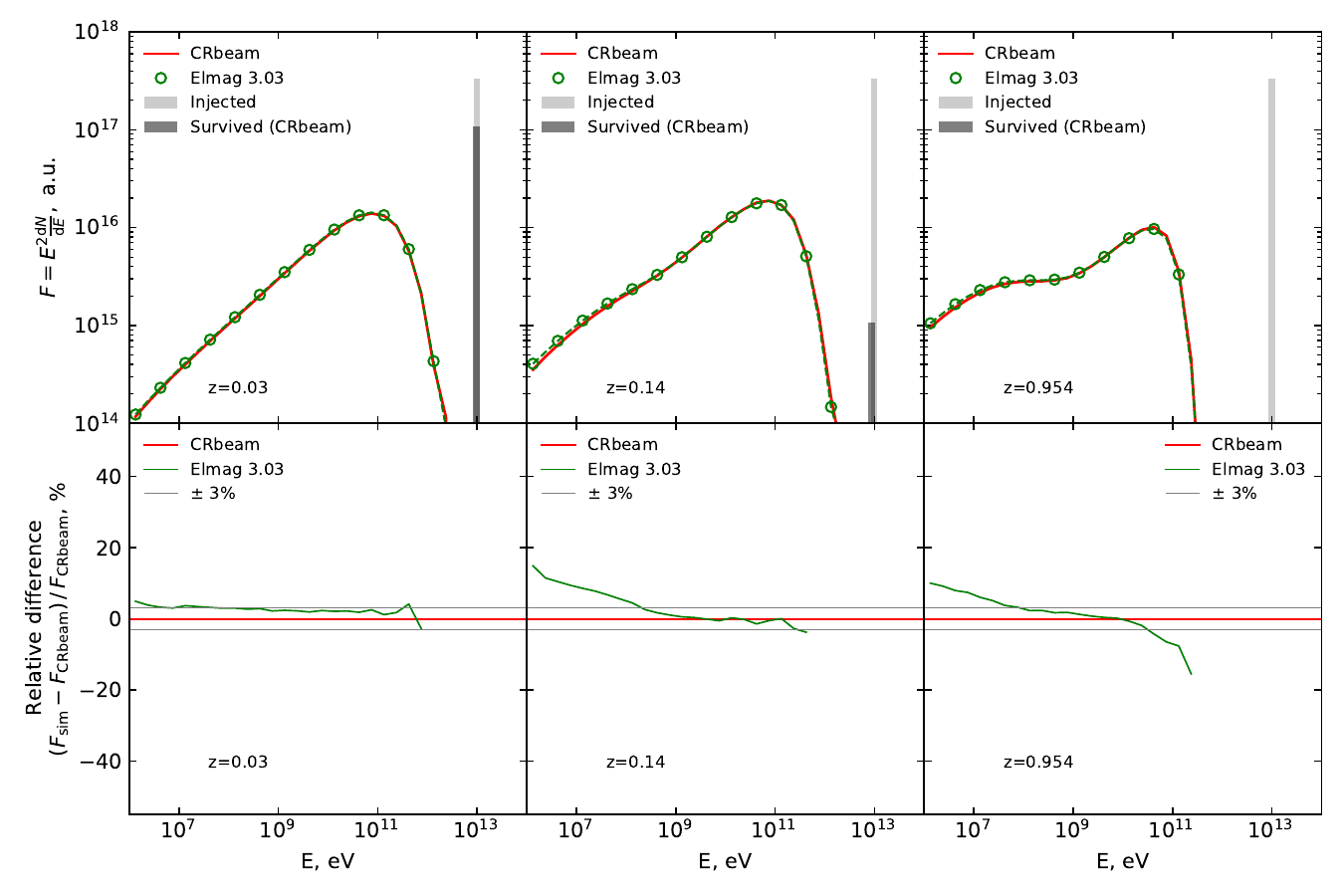}
    \caption{Model spectra of secondary \gr s produced by interaction of monoenergetic primary \gr s with energy $E_{\gamma_0}=10$~TeV. Same as Fig.~\ref{fig:spec_mono10TeV}, but calculated with CRbeam and ELMAG~3.03. After corrections to the implementation of the EBL model of \citet{franceschini08} in ELMAG~3.03, CRbeam, and ELMAG show much better agreement. \label{fig:spec_mono10TeV_Elmag303}}
    \vspace{9cm}
\end{figure*}

\FloatBarrier

\section{Additional tests with different EBL models}\label{sec:appendixB}
\begin{figure*}[b]
    \begin{center}
        \includegraphics[width=0.66\linewidth]{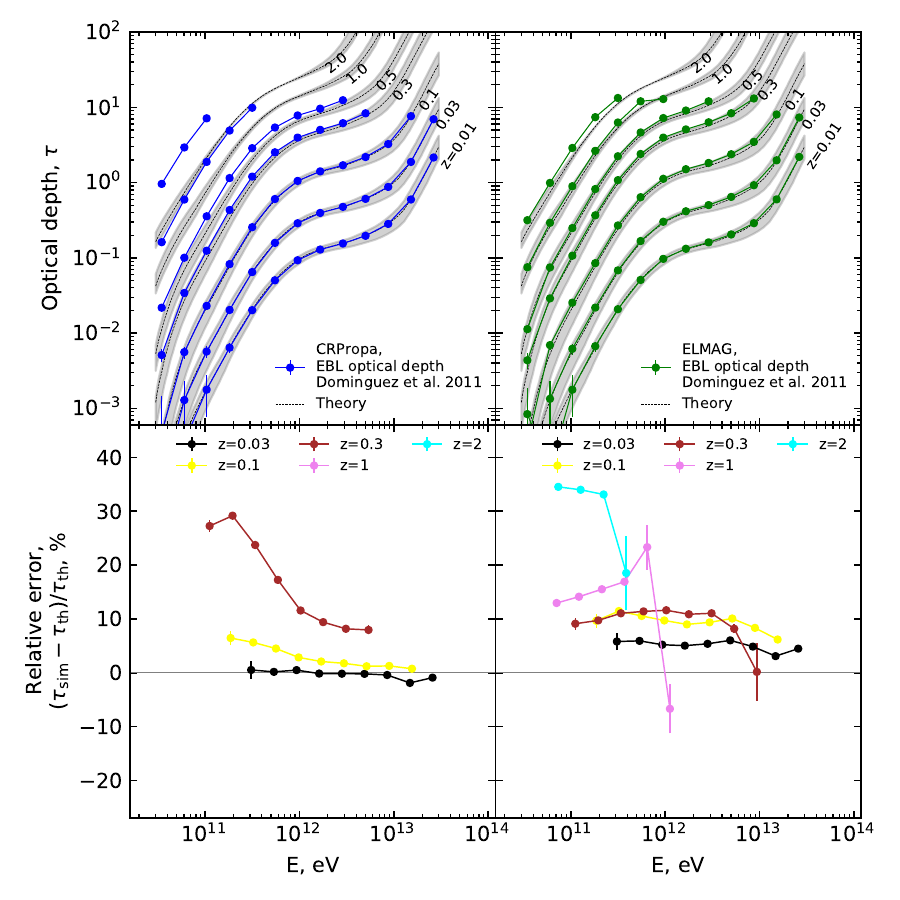}
        \caption{Optical depth of the EBL model of \citet{2011MNRAS.410.2556D} calculated with CRPropa and ELMAG in comparison with the analytical model. Figure layout and details are the same as in Fig.~\ref{fig:EBL_Franceschini08_opdep}. Grey bands indicate optical depth uncertainty provided by \citet{2011MNRAS.410.2556D}. Calculations with CRbeam are not shown as the EBL model of \citet{2011MNRAS.410.2556D} is not available in CRbeam.
        \label{fig:EBL_Dominguez2011}}
    \end{center}
    \vspace{8.5cm}
\end{figure*}
\begin{figure*}
    \includegraphics[width=\linewidth]{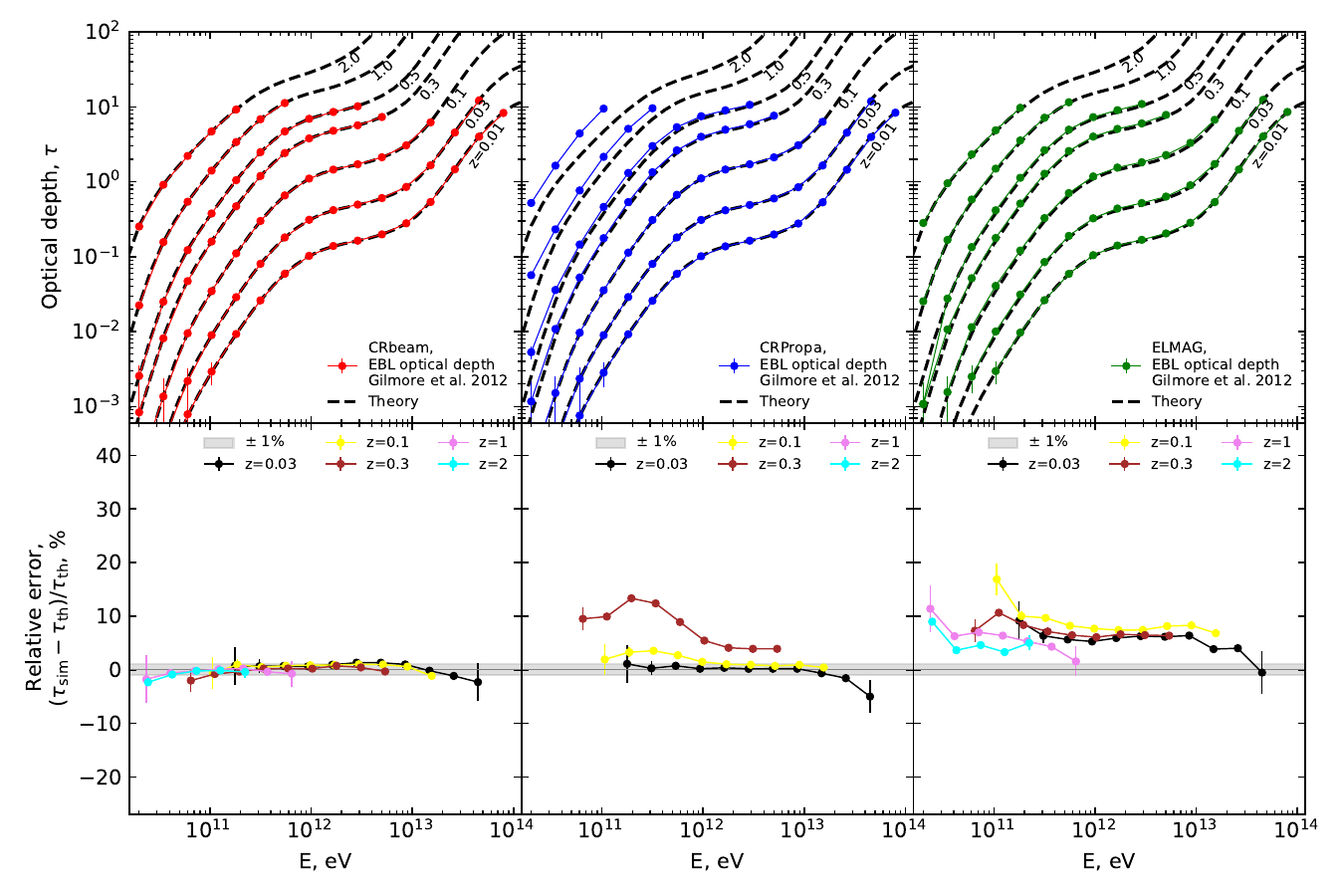}
    \caption{Optical depth of the EBL model of \citet{2012MNRAS.422.3189G} calculated with CRbeam, CRPropa, and ELMAG in comparison with the analytical model. Figure layout and details are the same as in Fig.~\ref{fig:EBL_Franceschini08_opdep}. \label{fig:EBL_Gilmore2012}}
\end{figure*}
\end{appendix}

\end{document}